\documentclass[aps,prd,preprint,nopreprintnumbers,nofootinbib,showpacs]{revtex4}

\usepackage{graphicx}
\usepackage{dcolumn}
\usepackage{bm}

\newcommand{\Uos}{\Upsilon(1S)}

\newcommand{\wNIS}{ISR}

\newcommand{\gtap}{\stackrel{>}{_\sim}}
\begin{document}

\preprint{CLNS 06/1970}       
\preprint{CLEO 06-15}         

\title{Improved Measurement of the Branching Fraction and Energy
Spectrum of  $\eta'$ from  $\Uos$ Decays}

\author{O.~Aquines}
\author{Z.~Li}
\author{A.~Lopez}
\author{S.~Mehrabyan}
\author{H.~Mendez}
\author{J.~Ramirez}
\affiliation{University of Puerto Rico, Mayaguez, Puerto Rico
00681}
\author{G.~S.~Huang}
\author{D.~H.~Miller}
\author{V.~Pavlunin}
\author{B.~Sanghi}
\author{I.~P.~J.~Shipsey}
\author{B.~Xin}
\affiliation{Purdue University, West Lafayette, Indiana 47907}
% ...

\collaboration{CLEO Collaboration}
\noaffiliation


\date{October 10, 2006}

\begin{abstract}
We present an improved  measurement of the $\eta'$ meson energy
spectrum  in $\Uos$ decays, using 1.2 $\mathrm{fb^{-1}}$ of data
taken at the $\Uos$ center-of-mass energy with the CLEO III
detector. We compare our results with models of the $\eta '$
gluonic form factor that have been suggested to explain the
unexpectedly large
 $B\to \eta 'X_s$ rate. Models based on perturbative QCD fail to fit
the data for large $\eta '$ energies, and thus an explanation
outside the realm of the Standard Model or an improved understanding
of non-perturbative QCD effects may be needed to account for this
large rate.

\end{abstract}

 \maketitle

\section{Introduction}
\begin{figure}[htbp]
\includegraphics*[width=2.8in]{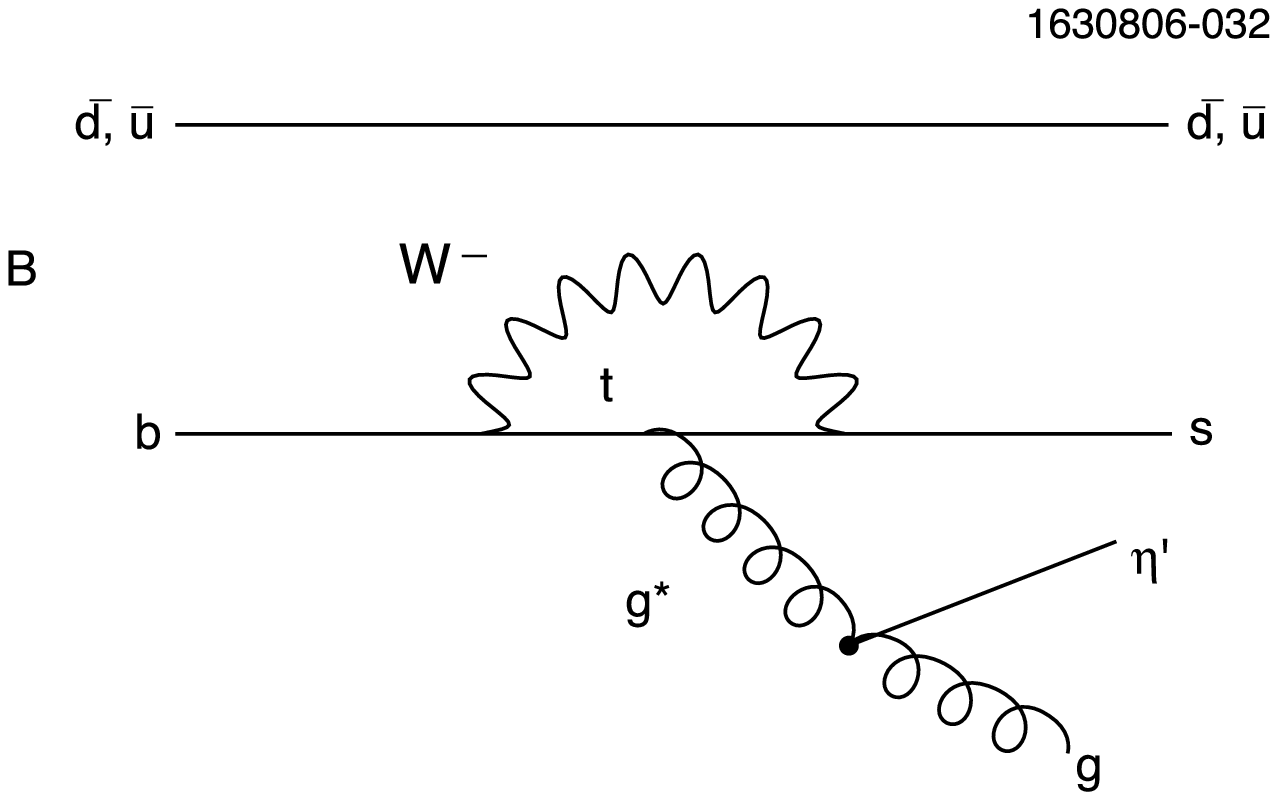}
\includegraphics*[width=3.0in]{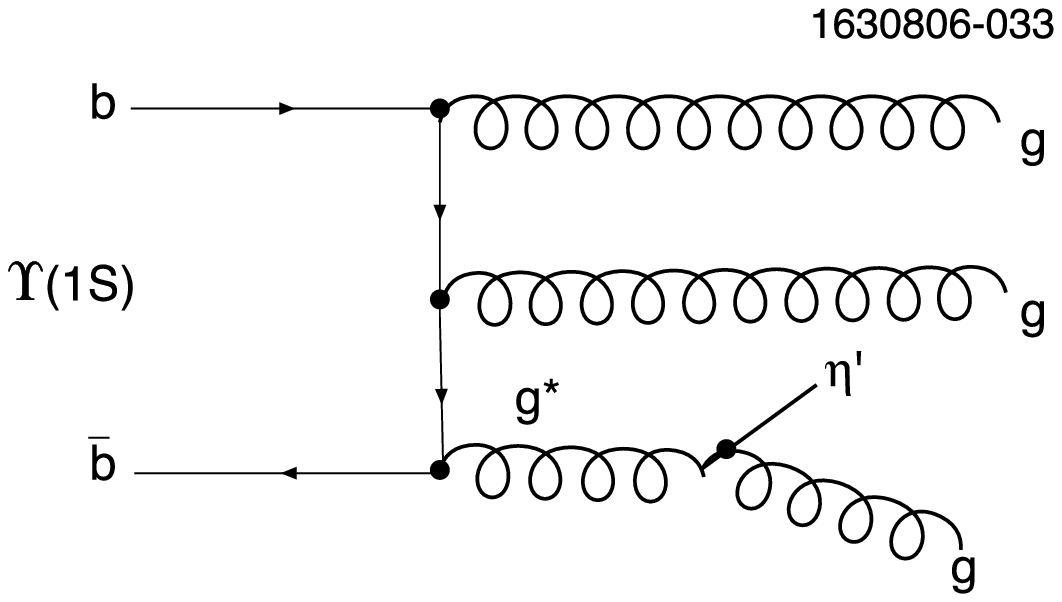}
\caption[A figure imported from a PostScript file] {Feynman
Diagram for $b\to s (g^\star \to g \eta')$ (left) and $\Uos\to
ggg^\star\to \eta ' X$ (right). }
    \label{etapvtx}
\end{figure}

CLEO observed a surprisingly large branching fraction for the
decay $B \to \eta' X_s$ at large momenta of the $\eta '$ meson
$p_{\eta '}:~{\cal B}(B \to \eta' X_s)|_{p_{\eta '}>2\ GeV}=
(6.2\pm 1.6\pm 1.3^{+0.0}_{-1.5})\times 10^{-4}$
\cite{Browder,Ernst}. BaBar \cite{babar} later obtained ${\cal
B}(B \to \eta' X_s)|_{p_{\eta '}>2\ GeV}= (3.9\pm 0.8 \pm 0.5 \pm
0.8)\times 10^{-4}$. This $\eta '$ momentum is beyond the end
point for most $b\to c$ decays, so the $\eta '$ yield from $b\to
c$ is expected to be only of the order of $1\times 10^{-4}$.
Predictions assuming factorization \cite{Kagan:1997,qcdfact}
estimate the charmless component of this branching fraction  to be
also about $1\times 10^{-4}$. Thus conventional calculations
cannot account for the measured rate and they also fail to predict
the right shape for the $\eta '$ momentum spectrum \cite{Kagan02}.
 These findings motivated intense theoretical
activity because new physics could account for such an
enhancement. However, Standard Model explanations have also been
proposed. For example, Atwood and Soni \cite{Atwood:1997bn} argued
that the observed excess is due to an enhanced $b\to s g$ penguin
diagram, complemented by a strong $\eta'gg^{\star}$ coupling,
induced by the gluonic content of the $\eta '$ wave function.
Fig.~\ref{etapvtx} (left) shows the corresponding Feynman diagram.
The high $q^2$ region of the $g^\star g \eta '$ vertex function
involved in this process  also affects fast $\eta'$ production in
$\Uos$ decay \cite{Kagan02,Atwood:1997bn,Ali:2000ci}, whose
relevant diagram is shown in Fig.~\ref{etapvtx} (right). Thus a
precise measurement of the $\eta '$ inclusive spectra from the
process $\Uos\to ggg^\star \to \eta' X$ can improve our
understanding of important $B$ meson decays.

The effective vertex function $\eta' gg^\star$ can be written as
\cite{Atwood:1997bn} $
H(q^2)\delta^{ab}\varepsilon_{\alpha\beta\mu\nu}q^{\alpha}k^{\beta}\varepsilon^{\mu}_{1}
\varepsilon^{\nu}_{2}$,  where $q$ is the $(g^\star)$ virtual
gluon's four-momentum, $k$ is the $(g)$ ``on-shell'' gluon's
momentum ($k^2=0$), $a$, $b$ represent color indices,
$\varepsilon^{\mu}_{1},\varepsilon^{\nu}_{2}$ are the polarization
vectors of the two gluons, and $H(q^2)$ is the $g^\star g\eta'$
transition form factor. Different assumptions on the form factor
dependence have been proposed \cite{Atwood:1997bn,Hou:1997wy,
Kagan:1997,Ali:2000ci,pQCD,kroll}.

While ARGUS was the first experiment to study the inclusive $\eta
'$ production at the $\Uos$ \cite{argus}, they did not have enough
data to separate $\Uos\to ggg^\star$ from the other components
discussed below. CLEO II \cite{jc} was the first experiment to
have sufficient statistics to measure inclusive $\eta '$
production from the subprocess $\Uos\to ggg^\star$. These data
ruled out a class of form factors characterized by a very weak
$q^2$ dependence \cite{Atwood:1997bn,Hou:1997wy}. Subsequently,
several theoretical calculations \cite{Ali:2000ci,pQCD,kroll}
derived the perturbative QCD form factors from models of the $\eta
'$ wave function. Attempts to use CLEO II data to constrain the
model parameters \cite{ali-extended} were not conclusive, due to
the limited statistics at the end point of the $\eta '$ spectrum.
Thus, it was difficult to establish whether neglecting higher
order terms in the perturbative expansion was appropriate
\cite{ali-extended}. An improved measurement, based on a
higher-statistics sample, is important to provide an improved
determination of the QCD parameters, and, consequently, a more
stringent test of the theory. This work reports a new measurement
of the inclusive $\eta '$ spectrum from the process $\Uos \to
ggg^\star \to \eta ' X$ based on the largest $\Uos$ sample
presently available, more than a factor of 11 greater than the
previous study \cite{jc}.

\section{Data sample and analysis method}
We use 1.2 $\mathrm{fb^{-1}}$ of CLEO III data recorded at the
$\Uos$ resonance, at 9.46 GeV center-of-mass energy, containing
$21.2\times 10^6$ events, and off-resonance continuum data collected
at center-of-mass energies of 10.54 GeV (2.3 $\mathrm{fb}^{-1}$).

The CLEO III detector includes a high-resolution charged particle
tracking system \cite{track}, a CsI electromagnetic calorimeter
\cite{csi}, and a Ring Imaging Cherenkov (RICH) hadron
identification system \cite{rich}. The CsI calorimeter measures
the photon energies with a resolution of 2.2\% at $E=1$ GeV and
5\% at $E$=100 MeV. The tracking system also provides charged
particle discrimination, through the measurement of the specific
ionization $dE/dx$.

We detect $\eta '$ mesons through the channel $\eta '\to\eta
\pi^+\pi^-$, with $\eta \to \gamma\gamma$. The branching fractions
for these processes are (44.5$\pm$1.4)\% and (39.38$\pm$0.26)\%
\cite{pdg06} respectively. We identify single photons based on
their shower shape. The photon four-vectors are constrained to
have invariant mass equal to the nominal $\eta$ mass.
Subsequently, $\eta$ candidates are combined with two oppositely
charged tracks to form an $\eta '$. Loose $\pi$ consistency
criteria based on $dE/dx$ measurements are applied to the charged
tracks.

The gluonic $\eta '$ production at the $\Uos$  is expected  to be
dominant only at very high $q^2$, or, equivalently, at high $\eta '$
scaled energy $Z$, where $Z$ is defined as
\begin{equation}
Z\equiv \frac{E_{\eta '}}{E_{\text{beam}}} =\frac{2E_{\eta
'}}{M_{\Uos}},
\end{equation}
where $E_{\eta '}$ is the $\eta '$ energy and $E_{\text{beam}}$ is
the beam energy. Enhanced $\eta '$ production at high $Z$ would
indicate a large $\eta 'g^{\star} g$ coupling.

\begin{figure}[htbp]
\includegraphics*[width=6.75in]{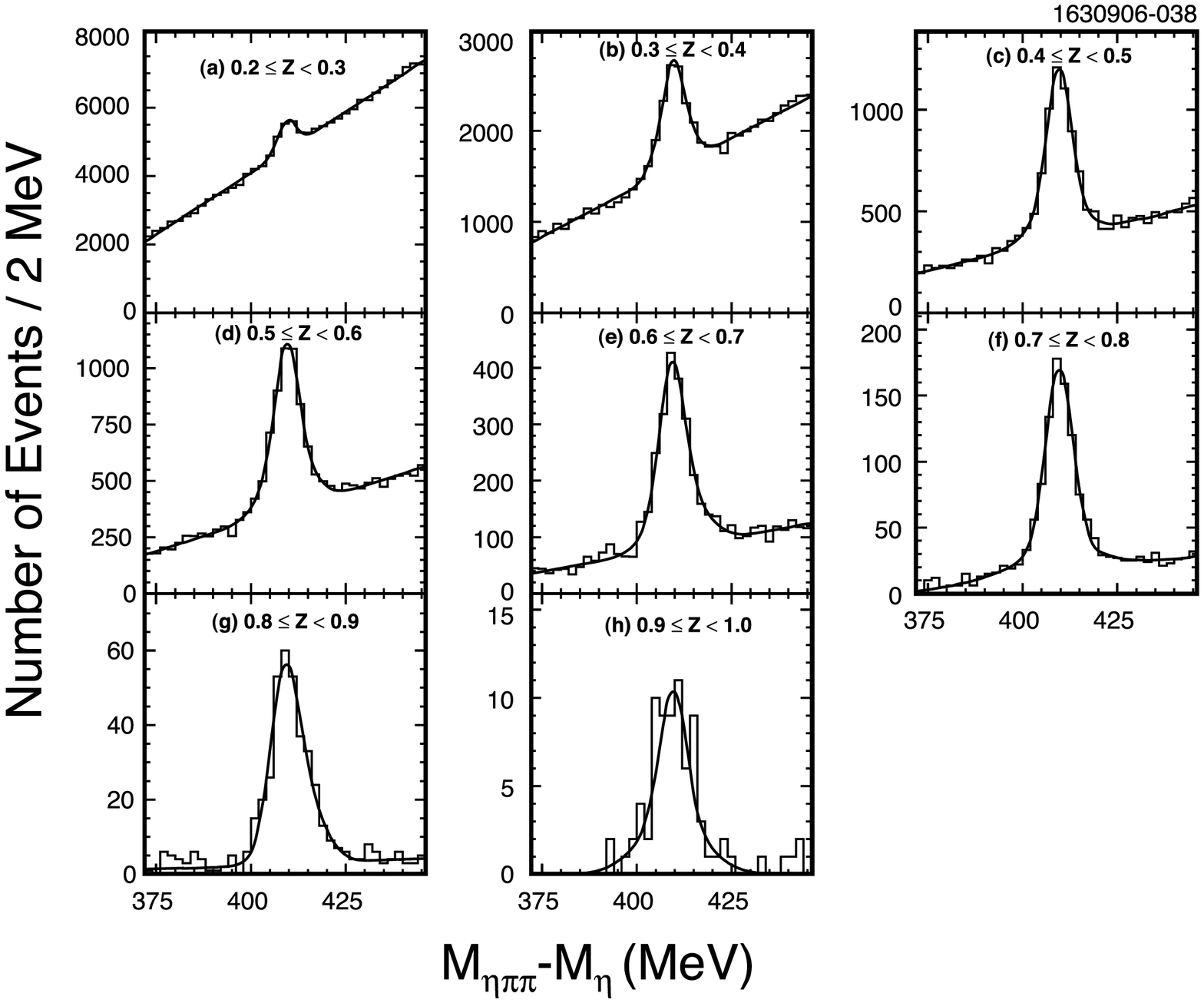}
\caption{\label{fig:xm1sz} The spectra of the difference of the
$\eta\pi^+\pi^-$
  and $\eta$ invariant masses in
  different $Z$ ranges reconstructed from $\Uos$ data,
  fit with a single Gaussian function for the signal and a first-order
  polynomial for the background.}
\end{figure}

\begin{figure}[htbp]
\includegraphics*[width=6.75in]{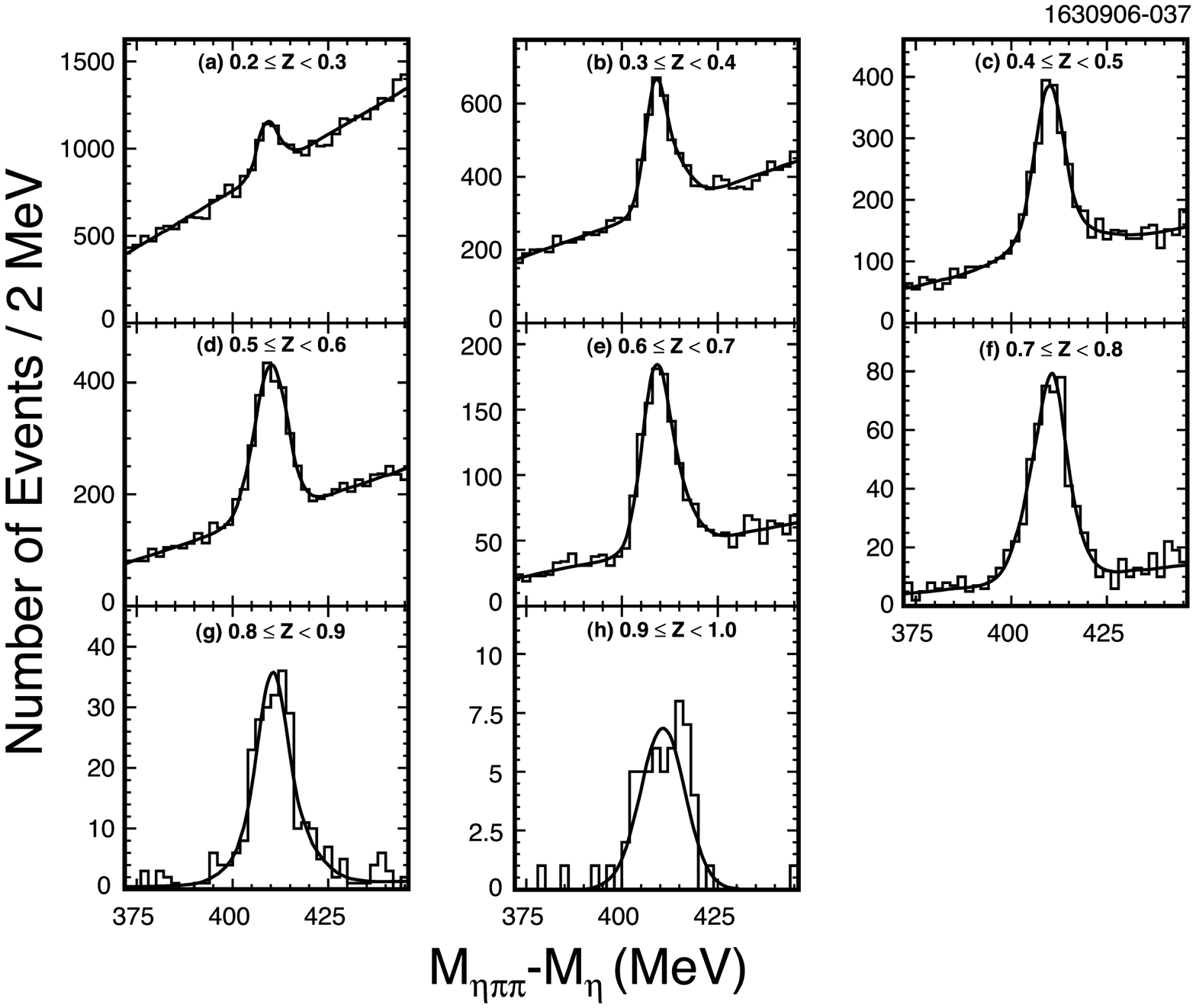}
\caption{\label{fig:xm4sz} The spectra of the difference of the
$\eta\pi^+\pi^-$
  and $\eta$ invariant masses in
  different $Z$ ranges reconstructed from continuum data taken at
  a center-of-mass energy of 10.54 GeV,
  fit with a single Gaussian function for the signal and a first order
  polynomial for the background.}
\end{figure}

For low-energy $\eta '$ mesons, photons coming from low energy $\pi
^0$s are a severe source of background. Thus a $\pi ^0$ veto is
applied for $Z < 0.5$, whereby photon pairs that have an invariant
mass consistent within 2.5 $\sigma$ with the nominal $\pi ^0$ mass
are not included as the candidate photons for $\eta$ reconstruction.
We consider only $\eta '$ with scaled energy $Z$ between 0.2 and 1
and divide this range into eight equal bins. Fig.~\ref{fig:xm1sz}
shows the $\eta '$ yields in these bins for the $\Uos$ sample.
Fig.~\ref{fig:xm4sz} shows the corresponding distributions from the
continuum sample taken at a center-of-mass energy of 10.54 GeV. In
order to derive the $\eta '$ signal yields,  we fit the $\Delta
M_{\eta '\eta}$ distributions [$\Delta M_{\eta '\eta}\equiv
M(\pi^+\pi^-\eta)-M(\eta)$] in each $Z$ bin with a Gaussian function
representing the signal, and a polynomial background. The Gaussian
is used only to define a $\pm 2.5 \sigma$ signal interval. Then the
$\eta '$ yield in this interval is evaluated counting events in the
signal window, after subtracting the background estimate obtained
from the fit function. As the $\Delta M_{\eta '\eta}$ signal is not
described well by a single Gaussian function, this procedure
minimizes systematic uncertainties associated with the choice of an
alternative signal shape.

Information on the gluon coupling of the $\eta '$ can be drawn only
from the decay chain $\Uos \to ggg^{\star}\to \eta ' X$, described
by the Feynman diagram in Figure 1. Thus we need to subtract both
continuum $\eta '$ production and $\eta '$ from the process $\Uos
\to \gamma ^\star\to q\bar{q}$. The latter component is estimated
using
\begin{equation}
{\cal B}(\Uos\to q\bar{q})= R\cdot {\cal B}(\Uos\to \mu ^+\mu^-)=
(8.83\pm 0.25)\%,
\end{equation}
where $R$ is the ratio between the hadronic cross section $\gamma
^\star \to q\bar{q}$ and the di-muon cross section $\gamma ^\star
\to \mu ^+\mu^-$ at an energy close to 9.46 GeV. We use $R=3.56\pm
0.07$ \cite{ammar:1998} and ${\cal B}(\Uos\to \mu ^+\mu^-)= (2.48
\pm 0.05)$\% \cite{pdg06}. The yield $N(\Uos \to ggg^\star )$ is
estimated with the relationship
\begin{equation} N(\Uos \to ggg^\star ) = N_{\mathrm{had}} - N(\gamma^\star\to q\bar{q}) -
N(\Uos\to q\bar{q}),\label{eq:ggg}
\end{equation}
where $N_{\mathrm{had}}$ is the number of hadronic events in our
sample, and $N(\gamma^\star\to q\bar{q})$ is the number of continuum
events derived from the 10.54 GeV continuum data set, corrected for
the luminosity difference between resonance and continuum data, and
the center-of-mass dependence of the cross section for the process
$\gamma ^\star \to q\bar{q}$.

\begin{figure}[htbp]
\includegraphics*[width=3.9in]{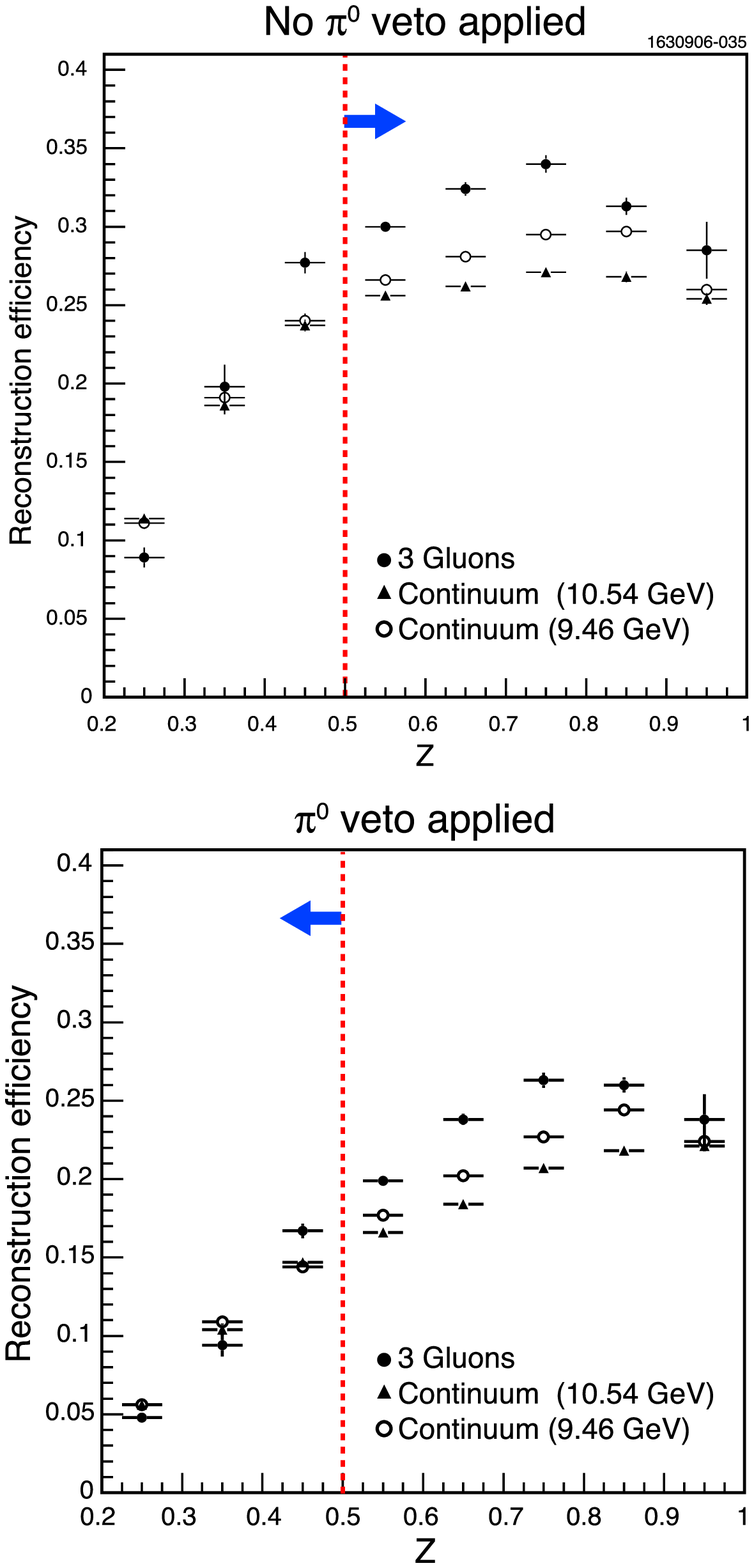}
\caption{\label{fig:eff}The $\eta'$ reconstruction efficiencies as
function of $Z$
 for different MC samples with no $\pi^0$ veto, and with $\pi^0$ veto in
 photon selection. The $\pi^0$ veto was applied in this analysis for $Z<0.5$.}
\end{figure}

The two dominant components of the $\eta '$ spectrum have different
topologies: $\Uos \to ggg^\star$ produces a spherical event
topology, whereas $q\bar{q}$ processes are more jet-like. This
difference affects the corresponding reconstruction efficiencies.
The $\gamma gg$ cross section is only about 3\% of the $ggg^\star$
cross section; thus we make no attempt to subtract the former
component from the latter. Fig.~\ref{fig:eff} shows the efficiencies
obtained for the two event topologies with CLEO III Monte Carlo
studies. We use GEANT-based \cite{geant} Monte Carlo samples,
including $\Uos$ and continuum samples. In order to use the
continuum sample taken at 10.54 GeV center-of-mass energy for
background subtraction, we develop a ``mapping function" to correct
for the difference in phase space and $Z$ range spanned in the two
samples. The procedure is described in detail in Ref.~\cite{jc}. In
brief, we use the $\eta '$ energy distribution functions  for the
Monte Carlo continuum samples at center-of-mass energies equal to
9.46 and 10.54 GeV and obtain a relationship between the measured
$Z$ at higher center-of-mass energy $Z_{10.54}$:
\begin{equation}\label{eq:Z}
Z_{9.46} = -0.215 \times 10^{-2} + 1.2238\ Z_{10.54} - 0.6879\
(Z_{10.54})^2
               + 0.8277\ (Z_{10.54})^3 - 0.3606\ (Z_{10.54})^4\ . \\
\end{equation}

\begin{figure}[htbp]
\includegraphics*[width=6.5in]{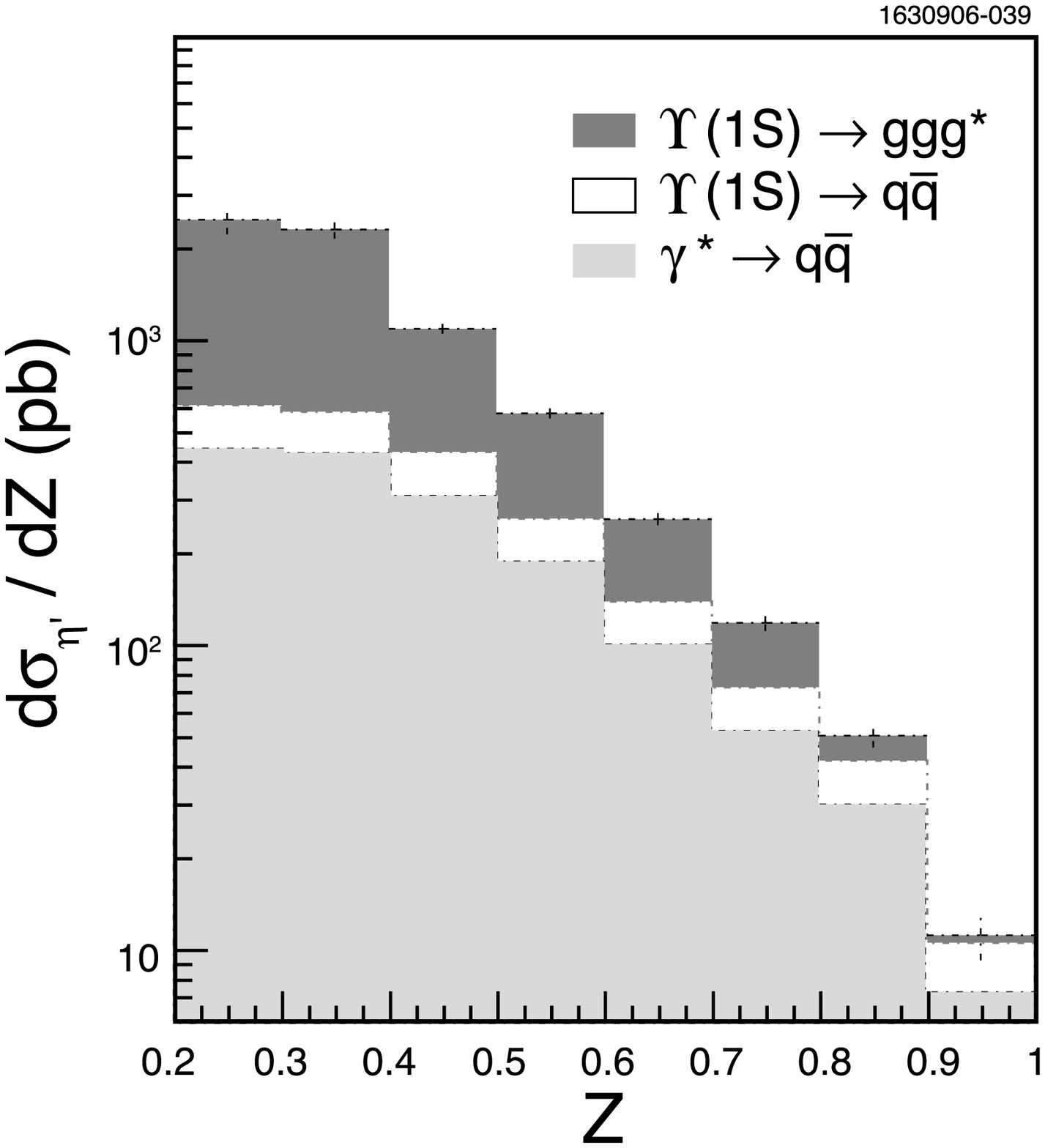}
\caption{\label{fig:etapcross}The measured differential cross
    sections $d\sigma_{\eta '}/dZ$ for a) $\gamma ^\star\to q\bar{q}\to\eta'X$
    (shaded), b) $\Uos\to q\bar{q}\to\eta'X$ (white) and c) $\Uos\to
    ggg^\star\to\eta'X$ (black).}
\end{figure}

We  derive the $\eta '$ spectrum from the decay $\Uos\to
ggg^\star\to \eta 'X$ from the efficiency corrected $\eta '$ yields
in each $Z$ bin, subtracting the contributions from continuum  and
$\Uos \to q\bar{q}$ as shown in Equation \ref{eq:ggg}. In this case,
the $\eta '$ spectrum from the process $\Uos \to q\bar{q}\to\eta' X$
is corrected for the distortion introduced by initial state
radiation (ISR) on the continuum $\eta '$ energy spectrum used to
account for this effect.  The partial yield $N(\Uos \to
q\bar{q})(Z)$ is estimated with the relationship
\begin{eqnarray}
N(\Uos \to q\bar{q})(Z) & = & N(\gamma^\star\to q\bar{q}\to\eta' X)(Z)\times R_{\mathrm{\wNIS}} \times \frac{\sigma_{\Uos\to q\bar{q}}}{\sigma_{e^{+}e^{-}\to q\bar{q}}} \nonumber \\
            & = & N(\gamma^\star\to q\bar{q}\to\eta' X)(Z)\times R_{\mathrm{\wNIS}} \times \frac{\sigma_{\Uos\to\mu^{+}\mu^{-}}}{
\sigma_{e^{+}e^{-}\to\mu^{+}\mu^{-}}}, \label{eq:fqq}
\end{eqnarray}
where $R_{\mathrm{\wNIS}}$ accounts for the difference between the
$\Uos\to q\bar{q}\to \eta ' X$ and the $\gamma ^\star\to q\bar{q}\to
\eta ' X$ spectra due to initial state radiation (ISR) effects,
estimated using Monte Carlo continuum samples with and without ISR
simulation, and $\sigma(\Uos\to q\bar{q})/\sigma(e^{+}e^{-}\to
q\bar{q})$ accounts for the relative cross section of these two
processes. The correction factor $R_{\mathrm{\wNIS}}$ differs from 1
by a few percent at low $Z$ and as much as 25\% at the end point of
the $\eta '$ energy. The cross sections used are
$\sigma(\Uos\to\mu^{+}\mu^{-})=0.502\pm 0.010 $ nb \cite{pdg06} and
$\sigma(e^{+}e^{-}\to\mu^{+}\mu^{-})=1.372\pm 0.014 $ nb
\cite{fpair}. Fig.~\ref{fig:etapcross} shows the measured
differential cross sections $d\sigma_{\eta '}/dZ$ for the processes
$\Uos\to ggg^\star$, $\Uos \to q\bar{q}$, and, $\gamma ^\star \to
q\bar{q}$.

\begin{figure}[htbp]
\includegraphics*[width=7in]{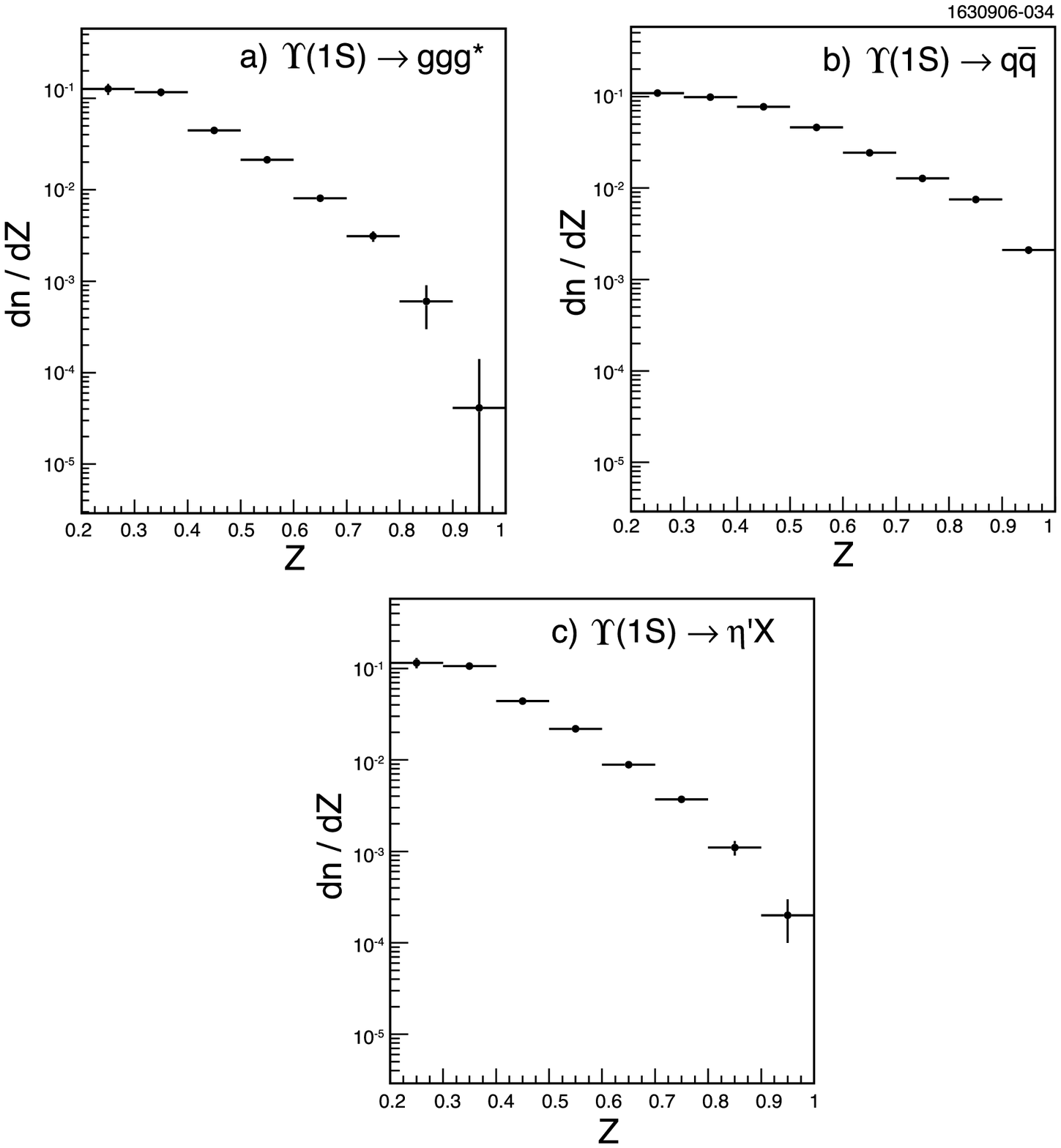}
\caption{\label{fig:zspect}The energy distribution function
$dn/dZ$ as defined in
  the text for a) $\Uos\to ggg^\star\to\eta'X$, b) $\Uos\to q\bar{q}\to\eta'X$, and c) $\Uos\to\eta'X$.}
\end{figure}

Theoretical predictions give the energy distribution function
$dn/dZ\equiv [1/N(\Uos \to ggg^\star)]\times dN(\Uos\to ggg^\star
\to \eta' X)/dZ$; we obtain the corresponding experimental quantity
by dividing by the total number of $N(\Uos \to ggg^\star)$,
estimated by applying Equation ~\ref{eq:ggg} without any $Z$
restriction. Figure~\ref{fig:zspect}.a) shows the $\Uos \to
ggg^\star\to \eta ' X$ energy distribution function, whereas
Fig.~\ref{fig:zspect}.b) and c) show the corresponding distributions
for $\Uos \to q\bar{q}\to \eta ' X$, normalized with respect of the
total number of $\Uos\to q\bar{q}$ and $\Uos \to \eta ' X$,
normalized with respect to the total number of $\Uos$.

The inclusive $\eta '$ production at the $\Uos$ is expected to be
dominated by the transition $\Uos\to ggg^\star\to \eta 'X$ only at
high $\eta '$ energy. The energy at which this occurs cannot be
predicted from first principles: an empirical criterion is the
$\chi ^2$ of the theory fit to the data. For example, a numerical
analysis of the CLEO II data \cite{ali-extended} obtained a $\chi
^2$ of 2.4 for three degrees of freedom, using the 3 experimental
points at $Z\ge 0.7$, and $\approx$ 24 for 4 degrees of freedom
using the 4 points at $Z\ge 0.6$. This observation led Ali and
Parkhomenko to conclude that the $Z$ region likely to be dominated
by $\Uos\to ggg ^\star\to \eta ' X$ starts at $Z=0.7$. Thus we
quote global branching fractions for $\Uos\to \eta 'X$ and the
corresponding results for $Z\ge 0.7$ separately.

Table~\ref{tab:error} summarizes the dominant components of the
systematic uncertainties. The overall relative errors on the $\eta
'$ branching fractions are $\pm 8.1\%$ for $q\bar{q}\to \eta ' X$,
$\pm 9.1\%$ for $ggg^{\star} \to \eta ' X$ for $Z>0.7$ and $\pm
7.2\%$ for all other branching fractions.

\begin{table}[ht]
\begin{center}
\begin{tabular}{|l|c|c|c|}
\hline \hline Sources     & $ggg^\star$ Sample ($Z>0.7$) &
$q\bar{q}$ Sample & All others \\ \hline
Reconstruction efficiency of $\pi^\pm$  & 2.0 & 2.0 & 2.0\\
Reconstruction efficiency of $\eta $     &   5.0 &   5.0 &   5.0 \\
Number of $\eta'$ from fit              &   1.0 &   1.0 &   1.0 \\
Total number of $\Uos$                  & 1.0 & 1.0 & 1.0 \\
${\cal B}(\eta'\to\pi^+\pi^-\eta)$      & 3.4 & 3.4 & 3.4 \\ \hline
${\cal B}(\Uos\to q\bar{q})$            &  -  & 3.0 & -   \\
Ratio of integrated luminosity         & 1.9 &  1.0  & -   \\
$\sigma_{\Uos\to\mu^+\mu^-},\sigma_{e^{+}e^{-}\to\mu^{+}\mu^{-}}$            & 0.7 &  1.3  & -   \\
$\pi^0$ veto                            &  -  & 1.7 & 0.4 \\
$Z$ mapping                               &   6.0 &  3.0  & 3.0   \\
\hline
Total                                   &  9.1 & 8.1  & 7.2 \\
\hline\hline
\end{tabular}
\caption{\label{tab:error} The components of the systematic errors
(\%) affecting the branching fractions reported in this paper.}
\end{center}
\end{table}

Thus we obtain
\begin{eqnarray}
n(\Uos\to (ggg^\star) \to\eta'X)\equiv \frac{N( \Uos\to ggg^\star
           \to\eta'X)}{N(\Uos\to ggg^\star)}
           &=& (3.2 \pm 0.2 \pm 0.2) \%, \nonumber\\
n(\Uos\to(q\bar{q})\to\eta'X)\equiv \frac{ N(\Uos\to
           q\bar{q}\to\eta'X)}{N(\Uos\to q\bar{q})}
           &=& (3.8 \pm 0.2 \pm 0.3)\%, \nonumber\\
n(\Uos\to\eta'X)\equiv \frac{ N(\Uos\to\eta'X)}{N(\Uos)}
           &=& (3.0 \pm 0.2 \pm 0.2) \%. \label{eq:br}
\end{eqnarray}

The $\Uos\to\eta'X$ branching fractions at high momentum ($\rm
Z>0.7$) are measured to be
\begin{eqnarray}
n(\Uos\to(ggg^\star)\to\eta'X)_{Z>0.7}
           &=& (3.7\pm 0.5 \pm 0.3)\times 10^{-4}, \nonumber\\
n(\Uos\to(q\bar{q})\to\eta'X)_{Z>0.7}
           &=& (22.5\pm 1.2 \pm 1.8)\times 10^{-4}, \nonumber\\
n(\Uos\to\eta'X)_{Z>0.7}
           &=& (5.1 \pm 0.4 \pm 0.4)\times 10^{-4}.\label{eq:br7}
\end{eqnarray}

\section{Comparison with theory and conclusions}
A. Kagan \cite{Kagan02} used the ratio $R_{Z>0.7}$, defined as
\begin{equation}
R_{Z>0.7}\equiv\left[\frac{n_{\mathrm{th}}}{n_{\mathrm{exp}}}\right]_{Z>0.7},
\end{equation}
to obtain a first rough discrimination between form factors having
drastically different $q^2$ dependence. At the time that
Ref.~\cite{Kagan02} was published, the comparison was based on 90\%
C.L. upper limits on the data. This test repeated with our present
data give values of $R_{Z>0.7} \gtap 74$ for a representative slowly
falling form factor \cite{Hou:1997wy}, $R_{Z>0.7} \gtap 25$ for the
intermediate form factor studied by Ref.~\cite{Kagan:1997}, and
$R_{Z>0.7} \gtap 2$ for the perturbative QCD inspired shape. Thus
the last shape is the closest to the data, but it does not match
them very well.

\begin{figure}[htbp]
\includegraphics*[width=3.5in]{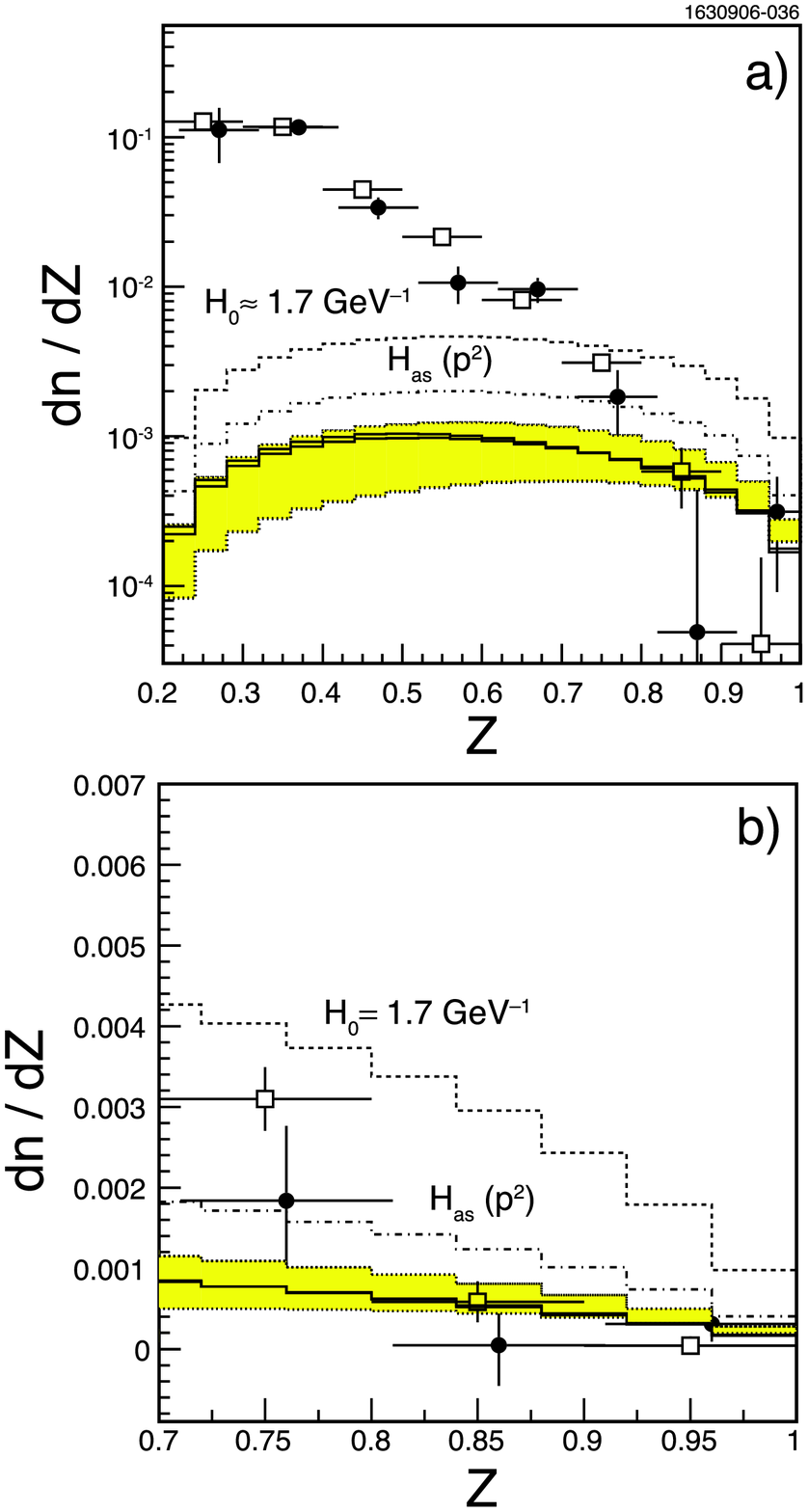}
\caption[A figure imported from a PostScript file]{
    Energy spectrum of the $\eta '$-meson in the decay
    $\Uos\to\eta'X$ (open squares correspond to  the data
    presented in this paper, filled circles are the previously
    reported CLEO II data \cite{jc}): a) measured spectra (log
    scale);
    b) expanded view of the $Z\ge 0.7$ region
    to show the comparison with the theoretical
    predictions more clearly (linear scale). The dashed curve corresponds to a constant
    value of the function $H(p^{2})=H_{0}\simeq 1.7$ GeV$^{-1}$, and the
    dash-dotted curve ($H_{\mathrm{as}}(p^{2})$) corresponds to the asymptotic form
    of the $\eta '$-meson LCDA \cite{ali-extended} (i.e., $B_{2}^{(q)}=0$ and
    $B_{2}^{(g)}=0$). The spectrum with the Gegenbauer coefficients \cite{ali-extended} in
    the combined best-fit range of these parameters is shown in the
    shaded region. The solid curve
    corresponds to the best-fit values of the parameters in the
    form factors from Ref. \cite{ali-extended}
    from the analysis of the $\Uos\to\eta'X$ CLEO II data alone.
    }
    \label{fig:cleocomp}
\end{figure}

Several perturbative QCD calculations of this process exist, and are
based on different choices of the form factor $H(q^2)$: Kagan and
Petrov \cite{Kagan02} assume $H(q^2)\approx\ \text{const} \approx
1.7\ \rm{GeV}^{-1}$; Ali and Parkhomenko relate $H(q^2)$ to the
expansion of the two light-cone distribution amplitudes (LCDA)
describing the quark and gluon components of the $\eta '$ wave
function \cite{ali-extended}. Figure~\ref{fig:cleocomp} shows the
measured $dn/dZ$ distribution, compared with three representative
choices for $H(q^2)$: $H(q^2)=H_0=1.7$ GeV$^{-1}$,
$H_{\mathrm{as}}$, based on the asymptotic form of the $\eta '$
meson LCDAs, and $H(q^2)$ corresponding to the Ali and Parkhomenko
\cite{ali-extended} formalism, with the parameters extracted from
the previous CLEO II data and the constraints from the $\eta
'-\gamma$ transitions \cite{kroll}. Note that in the range where
$\Uos \to ggg^\star$ is the relevant source of $\eta '$ most of the
discrepancy between theory and data occurs in the $Z=0.7$ bin. In
fact, the $\chi ^2$ for the fit of the new data with this
theoretical parametrization is 27 for 3 degrees of freedom. This may
imply that higher order terms in the QCD expansion need to be taken
into account, or that the $\Uos \to ggg^\star$ is not the dominant
source of $\eta '$, at least at a scaled energy as high as $Z=0.7$.

In conclusion we have measured the energy spectra of the $\eta '$
meson in the decay $\Uos \to \eta ' X$. Our results are not very
well described by existing models based on strong gluonic coupling
of the $\eta '$. Thus the observed $B\to \eta ' X$ inclusive
branching fraction is unlikely to be explained by an enhanced
$g^\star g \eta '$ form factor, and an explanation outside the
realm of the Standard Model or an improved understanding of
non-perturbative QCD effects may be needed to account for this
large rate.

\section{Acknowledgements}
We would like to thank A. Kagan and A. Ali for useful discussions
and for providing us with their calculations. We gratefully
acknowledge the effort of the CESR staff in providing us with
excellent luminosity and running conditions. D.~Cronin-Hennessy and
A.~Ryd thank the A.P.~Sloan Foundation. This work was supported by
the National Science Foundation, the U.S. Department of Energy, and
the Natural Sciences and Engineering Research Council of Canada.

\end{document}